\documentclass[aps,pra,reprint, superscriptaddress]{revtex4-1}
\usepackage[utf8]{inputenc}

\usepackage{epsfig,graphics,graphicx}
\usepackage{dcolumn}
\usepackage{bm}
\usepackage{color}
\usepackage{amsmath,amssymb,amsthm}
\usepackage{float}
\usepackage{centernot}
\usepackage{bbold}
\usepackage{amsfonts}
\usepackage{mathtools}
\usepackage{fullpage}
\usepackage{units}
\usepackage[usenames,dvipsnames]{xcolor}
\usepackage{braket}
\usepackage{indentfirst}

\usepackage[T1]{fontenc}

\usepackage{mathpazo}
\usepackage{dsfont}

\usepackage{mathptmx}
\usepackage{amssymb}
\usepackage{amsmath}
\usepackage{latexsym}
\usepackage{amsfonts}

\usepackage{hyperref}
\hypersetup{pdfpagemode=UseNone}

\newtheorem{theorem}{Theorem}
\newtheorem{definition}[theorem]{Definition}

\newtheorem{lemma}[theorem]{Lemma}
\newtheorem{proposition}[theorem]{Proposition}

\newcounter{rem}
\newcommand{\mc}[1]{\mathcal{#1}}

\def\>{\rangle}
\def\<{\langle}

\renewcommand{\rho}{\varrho}

\def\textbf#1{{\bf #1}}

\def\beq{\begin{equation}}
\def\eeq{\end{equation}}
\def\beqa{\begin{eqnarray}}
\def\eeqa{\end{eqnarray}}
\def\eea{\end{array}}
\def\bea{\begin{array}}
\newcommand{\bei}{\begin{itemize}}
\newcommand{\eei}{\end{itemize}}
\newcommand{\bee}{\begin{enumerate}}
\newcommand{\eee}{\end{enumerate}}
\def\bep{\begin{proposition}}
\def\eep{\end{proposition}}
\def\bel{\begin{lemma}}
\def\eel{\end{lemma}}
\def\bet{\begin{theorem}}
\def\eet{\end{theorem}}
\def\bed{\begin{definition}}
\def\eed{\end{definition}}
\usepackage{soul}
	
\begin{document}
	
\title{Coarse Graining of Partitioned Cellular Automata}

\author{Pedro C.S. Costa}
	
\affiliation{Department of Physics and Astronomy, Macquarie University,  \\  Sydney, New South Wales 2109, Australia }

\author{Fernando de Melo}
	
\affiliation{Centro Brasileiro de Pesquisas Físicas - CBPF -- Rua Dr. Xavier Sigaud 150, \\ Urca,  Rio de Janeiro, 22290-180, RJ, Brasil}
	
\date{\today}

\begin{abstract}
Partitioned cellular automata are known to be an useful tool to simulate linear and nonlinear problems in physics, specially because they allow for a straightforward way to define conserved quantities and reversible dynamics. Here we show how to construct a local coarse graining description of partitioned cellular automata. By making use of this tool we investigate the effective dynamics in this model of computation. All examples explored are in the scenario of lattice gases, so that the information lost after the coarse graining is related to the number of particles. It becomes apparent how difficult it is to remain with a deterministic dynamics after coarse graining. Several examples are shown where an effective stochastic dynamics is obtained after a deterministic dynamics is coarse grained. These results suggest why random processes are so common in nature.

\end{abstract}


\maketitle

\section{Introduction}

The process of emergence in physics typically occurs when we move from a microscopic to a macroscopic description~\cite{Kadanoff,Israeli,Oleg}. Frequently, because of the weak sensibility of our detectors, associated with the lack of information about the complete system, the dynamics we observe does not unveil knowledge about the totality of the microscopic system.
For instance, an electrically neutral structure, in general, is established out of interactions between positive and negative charges. Often our detector cannot access the full description of the system, and as such it gives us the information that the system is neutral. The same idea can be transposed to spin particles. Very often, our detectors cannot distinguish whether there are two neighboring particles with spins pointing to the same direction, and in the end, it only processes the information about an effective spin. But this is exactly what we want in several cases, that is, to work with less degrees of freedom, thus demanding fewer resources, while still catching all essential information.  In more general aspects, emergent processes arise spontaneously because of the high number of interacting subsystems, with no central control~\cite{choi2001supply}. Furthermore, even if we have the complete understanding of these individual parts, we cannot predict when and what will emerge, which makes the study of emergence a hard task ~\cite{Kadanoff}. 

In physics, a tool that is very often used to study emergence is known as coarse graining (CG). In statistical mechanics, the concept of CG appears when we deal with renormalization methods \cite{Kadanoff}, and it also plays an important role in models for biomolecular dynamics \cite{CG-bio}. Moreover, when a huge number of particles are considered in a microscopic system, one has to deal with several coupled differential equations. In general, in realistic cases, there are several boundary conditions involved in these problems, so that, in the end, one is forced to rely on numerical methods for differentials equation to describe systems with large number of degrees of freedom, which is very difficult to manage~\cite{PdeCA}. 

Then, in this particular case, a good alternative is to try to figure out which are the relevant degrees of freedom
to describe this system, i.e. the properties of interest in the simulation at issue. By doing that, fewer parameters can be employed, and it renders a more efficient simulation in terms of the required resources. Therefore, it is clear why it is so important in physics to understand and predict the emergence of large scale behavior in a system, starting from its microscopic description.

In the present work, cellular automata are employed to study emergence. A cellular automaton (CA) is a lattice of cells such that, at any moment in time, each cell is in one out of a finite set of discrete states. At each
discrete time step the state of each and every cell is updated according to some \textit{local transition function}. Cellular automata are paradigmatic forms and models of complex systems, since their temporal dynamics is totally given by local operations, without any central control~\cite{Neumman,Wolfram}. Many systems in nature have these characteristics, such as ant colonies \cite{Ants} and  brains \cite{BrainComp}; after all, despite the fact that the individual components of these systems are relatively well understood, as also are the local interaction between them, it is hard, if not impossible many times, to predict what will emerge in terms of formation of complex colonies and brain functionalities. Although CAs have a simple formulation -- local rules uniformly acting on all cells in synchronous fashion --  their dynamics is extremely rich, which render them appealing to create computational models for a range of systems, as in in biology \cite{CA_bio}, cryptography \cite{CA_cripto} and fluid dynamics \cite{FHP}.

Since the focus here is to study emergence in physics, where the properties of conservation and reversibility play an important role, we employ a cellular automata class known as partitioned cellular automata (PCA). Although the notions of reversibility and/or conservation are present in the context of CAs \cite{Morita,PP,NumCons,Kari2018} these properties can be achieved more easily in the PCA or block automaton, as proposed by Toffoli and Margolus~\cite{Margolus} and further developed by Morita \cite{Morita}. By employing a PCA, the concepts of reversibility and conservation become straightforward.

\begin{figure}[ht]
	\noindent \includegraphics[scale=0.28]{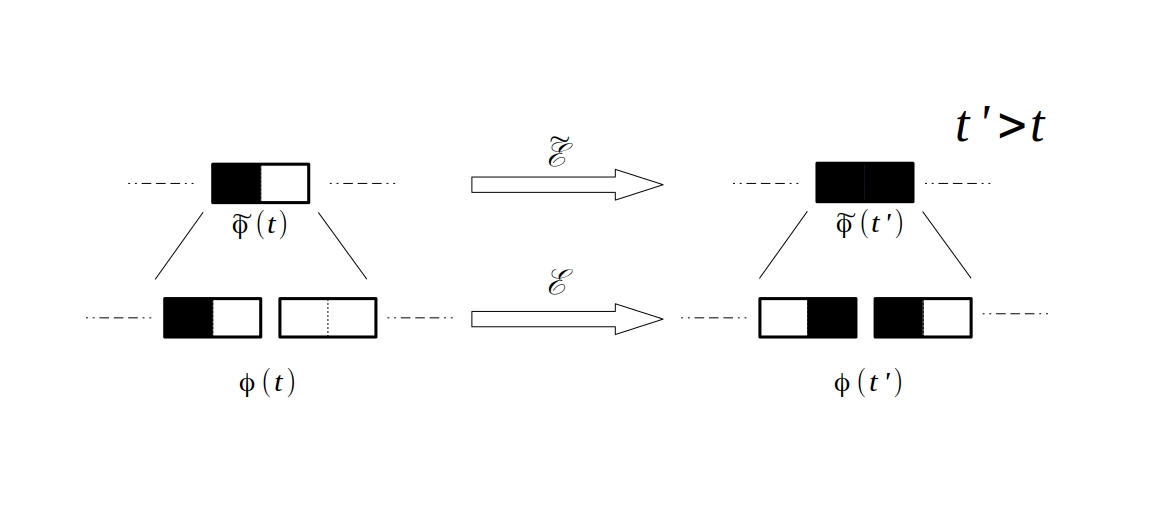}
	\caption{\small\label{fig:PCA_CGscheme}PCA coarse graining illustration. While $\mathcal{E}$ represents the transition function of some PCA, that maps its state $\phi$ from time $t$ to $t'$, $\mathcal{\tilde{E}}$ represents the effective transition function that now maps the PCA state $\tilde{\phi}$, achieved after the coarse graining, from $t$ to $t'$.} 
\end{figure}

In tune with the results by Israeli and Goldenfeld \cite{Israeli} and by Oleg \cite{Oleg}, our main goal is to develop a tool to study effective dynamics of classical systems. Just like they did, we developed a coarse graining technique in order to allow us to explore emergent dynamics in different scales. But there are two noticeable differences between our work and theirs. First while Oleg's work does not use any internal space structure, constrained with local rules of evolution and interaction, ours does, as it relies upon a description in terms of a PCA; see Fig. \ref{fig:PCA_CGscheme}. The second difference that departs our work from the one in \cite{Oleg} comes from the fact that, while he developed a classical CG technique to explore only stochastic processes our model can work both with deterministic and stochastic processes. In comparison with the results established in \cite{Israeli}, that employ the Wolfram CA's~\cite{Wolfram} and is more directed toward the computer science community, ours is more interesting for physics understanding, as PCAs can describe many distinct dynamics in physics~\cite{cellubook,FHP,HPP,CD98,cabio}. Furthermore the differences between our approach and that of \cite{Israeli} are also manifested in terms of their structural differences: as we rely upon a PCA, the structure of its transition function allows us to establish not only temporal but also spatial CG. All these differences will be clarified later on in the text.

This paper is organized as follows. In Section \ref{secI} we introduce a definition of partitioned cellular automata (PCAs), and also show their general behavior in one dimension in terms of permutation operators. In Section \ref{secII} the procedure for coarse graining the PCA is presented, which is then analysed and discussed in the subsequent section. Section \ref{secConc} concludes, by summarizing and commenting the results achieved, as well as discussing the perspectives of the procedure in possible future efforts.


 \section{Characterization of a PCA}\label{secI}
  
 Formally, a partitioned cellular automaton (PCA) can be defined as follows:
 \begin{definition}\label{def_PCA}
 	\text{[PCA]} A Partitioned Cellular Automaton is a 5-tuple $\left(L,\mathcal{N},\Sigma,\left\{ \mathcal{T}_{i}\right\} ,\left\{ \sigma_{i}\right\} \right)$ consisting of:
 	
 	\begin{enumerate}
 		\item A $d$-dimensional lattice of cells indexed by integers $L\subseteq\mathbb{Z}^{d}$;
 		
 		\item A finite neighborhood scheme $\mathcal{N}\subseteq L$;
 		
 		\item Each cell is divided in $n$ subcells,
 		and to the $i$-th subcell we assign a copy $\Sigma_{i}$ of a finite alphabet. The total alphabet associated to each cell is then  $\Xi=\Sigma_{0}\times\ldots\times\Sigma_{n-1}$;
 		
 		\item A finite set of $M$ tilings $\left\{ \mathcal{T}_{i}\right\}_{i=0}^{M-1} $. Each tiling is the union of identical non-overlapping tiles, $\mathcal{T}_i=\bigcup_j T_j^{(i)}$, with each tile  $T_j^{(i)}$ containing only subcells of neighboring cells;
 		
 		\item A set of local  functions $\left\{ \sigma_{i}\right\}_{i=0}^{M-1}$. The operator $\sigma_{i}$ is applied to each tile $T_j^{(i)}$  of the tiling $\mathcal{T}_i$.
 		\end{enumerate}
 \end{definition}
 
 With this definition, the transition function $\mathcal{E}:\Xi^{L}\rightarrow\Xi^{L}$, which updates the global automaton state $\varPhi_t\in \Sigma^{L}$  from the time $t$ to $t+1$, is given by 
 \begin{equation}
 \label{eq:transition_func}
 \mathcal{E}=\prod_{i=0}^{M-1}\left(\bigtimes_{T_{j}^{(i)}\in\mathcal{T}_{i}}\sigma_{i}\right).
 \end{equation}
 
 In this perspective, the state update, from $t$ to $t+1$, which is done by the transition function, can be divided by more than one set of local operators $\sigma$. The number of local operators is defined by the number of tilings, i.e. a uniform partition of the set of subcells, used to define the PCA.  This definition gives us freedom to access different dynamics and to apply our model to more complicated geometries. 
 
\begin{figure}[ht]
 	\noindent \includegraphics[scale=0.75]{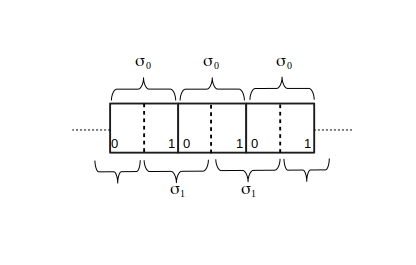}\caption{\small\label{fig:PCA_ex}Each cell is split into two subcells, and the operators $\sigma_{i}$ are applied in accordance with the two tilings.}
 \end{figure}

In order to work with tilings more precisely, it is convenient to put labels in each subcell.  Given the cell at position $x\in L$, its subcells are denoted by $x_i$, with $i\in \{0,\ldots,n-1\}$.  For instance, suppose we have a one-dimensional lattice, $L=\mathbb{Z}$, where each cell has two subcells, and the neighbor scheme is $\mathcal{N}_{x}=\left\{x-1,x,x+1\right\}$. In this case two tilings are sufficient to evolve the automaton: the first one given by $\mathcal{T}_0=\bigcup_{x\in \mathbb{Z}} T_x^{(0)}$, with each tile defined as $T_x^{(0)}=\{x_0,x_1\}$; the second tiling could be $\mathcal{T}_1=\bigcup_{x\in \mathbb{Z}} T_x^{(1)}$, each tile given by $T_x^{(1)}=\{x_1,(x+1)_0\}$. The first tiling is responsible for ``reading'' the state of each cell, while the second is responsible for the interaction between the neighboring cells. Now that the tilings' structure is established, the action of the operator is clear:
 \begin{eqnarray*}
 	\sigma_{0}:\left(\Sigma_{0}\right)_{x}\times\left(\Sigma_{1}\right)_{x}&\rightarrow&\left(\Sigma_{0}\right)_{x}\times\left(\Sigma_{1}\right)_{x}, 
	\\\sigma_{1}:\left(\Sigma_{1}\right)_{x}\times\left(\Sigma_{0}\right)_{x+1}&\rightarrow&\left(\Sigma_{1}\right)_{x}\times\left(\Sigma_{0}\right)_{x+1}
 \end{eqnarray*}
 for all $x\in \mathbb{Z}$. Therefore, in this example, the transition function can be written explicitly as 
 \begin{equation}
 \mathcal{E}=\left(\bigtimes_{T_{x}^{(1)}\in\mathcal{T}_{1}}\sigma_{1}\right)\left(\bigtimes_{T_{x}^{(0)}\in\mathcal{T}_{0}}\sigma_{0}\right).
 \end{equation}
 By choosing $\sigma_{i}$ for $i\in \{0,1\}$ as a permutation functions, which are reversible, the PCA becomes reversible. The sequence of steps leading to $\mathcal{E}$ in this example is illustrated in Fig. \ref{fig:PCA_ex}.
 
The operators $\sigma_i$ can be either deterministic or stochastic. In the first case, the local functions are given by permutation matrices $\pi^{(i)}$, while in the stochastic evolution it is given by a convex combination of permutations,   
\[
\sigma_i=\sum_{j=1}^{n!}p_{j}^i\pi^{(j)},
\]
with $p_{i}\geq0$ and $\sum_{i=1}^{n!}p_{i}=1$.

The restriction to employ only permutation operators naturally appears in  cases where the number of particles should be kept constant along the evolution. The description can be easily generalized to the case in which the number of particles is not conserved.


\section{Coarse graining the PCA} \label{secII}
Although the approach chosen here is quite general, in the sense that it can be applied to different geometries and with an arbitrary number of particles or excitations, for simplicity  we focus on the one-dimensional case with a single excitation. Given that, there is only the need to employ subcells with two states $\Sigma_i=\{0,1\}$, state 1 representing the particle existence (excitation), and state 0 the empty subcell. Thus, from now on, one bit per subcell is referred to herein as $\left(\mathbb{Z}_{2}\right)_{i}$, instead of $\Sigma_i$. Therefore, $\mathbb{Z}_{2}^{n}$ now stands for the finite set of cell states, given that there are $n$ bits per cell. Despite the fact that we restrict to one-dimensional PCAs, the cases with more than two subcells per cell are explored.  Interactions, however, will
remain only between two subcells from different cells, which means the tiles of the second tiling
have the structure $T_{x}^{(1)}=\left\{ x_{n-1},\left(x+1\right)_{0}\right\}$. Interaction between the cells
happens thus only across boundary subcells.   

Without loss of generality, the evolution will be restricted to the cases of two tilings, since with this number of tilings all non-trivial possible dynamics of the one-dimensional PCA can be accessed. Moreover, in order to allow interaction between the cells, we will always employ the Swap as the second operator, $\sigma_1=Swap$, the one related with the second tiling $\mathcal{T}_1$.

Since the present context relies on two tilings, with the maps $\sigma_0$ and $\sigma_1$ related to the first and the second tiling, respectively, we will often write $\mathcal{E}\left(\sigma_{1},\sigma_{0}\right)$, to indicate the transition function employed.

\subsection{The coarse graining procedure}

The first thing to be done in order to get the CG is to construct a supercell. The starting point is a PCA global  state at time $t$, $\varPhi_t$, with $\left|L\right|$ cells, each with $n$ subcells, 
\[
\varPhi_{t} \in \left(\mathbb{Z}_{2}^{n}\times\ldots\times\mathbb{Z}_{2}^{n}\right)_{\left|L\right|}.
\]
As the next step, $s$ cells are joined, $s$ being an integer defining the supercell size. Thus, a  PCA  global state in terms of supercells is achieved, 
\begin{equation}
\label{eq:SperCellPCA}
\varPhi_{t}^s \in \left(\mathbb{Z}_{2}^{sn}\times\ldots\times\mathbb{Z}_{2}^{sn}\right)_{{\left|L\right|}/s},
\end{equation}
with ${\left|L\right|}/s$ supercells. We need to stress the fact that the choice for ${\left|L\right|}$  is such that ${\left|L\right|}/s \in \mathbb{N}$. Furthermore, it is important to notice that while the number of cells is reduced when we move to the supercell representation, the number of subcells is increased in such a way that by the end of the process the total number of subcells is kept constant. By doing that, the same transition function is  still well defined in terms of  supercells. That is, at the present work we are considering the following equality $\mathcal{E}\varPhi_{t}^{s}=(\mathcal{E}\varPhi_{t})^s$. Once ${\left|L\right|}/s$ supercells are obtained, a CG map is constructed as follows:
\begin{equation}
\label{eq:first_part}
\Lambda_{CG}:\mathbb{Z}_{2}^{sn}\rightarrow\mathbb{Z}_{2}^{n'}.
\end{equation}
For being considered a coarse-graining map, we demand $n'<sn$. Some information about the full state is then lost after the action of $\Lambda_{CG}$.   Here we will be restricted to the case of $n'=n$. This map is applied to all supercells, in order to achieve a possible CA candidate with ${\left|L\right|}/s$ cells and with $n$ subcells,
\begin{equation}
\label{eq:low_to_up}
\varLambda_{CG}^{{\left|L\right|}/s}\varPhi_{t}^{s}=\tilde{\varPhi}_{T},
\end{equation}
where 
\[
\underset{\left|L\right|/s\text{ times}}{\Lambda_{CG}^{\left|L\right|/s}=\underbrace{\Lambda_{CG}\times\ldots\times\Lambda_{CG}}},
\]
and $\tilde{\varPhi}_{T}$ is a PCA global  state at time $T$ where in general $T\neq t$, as we will see later,  in the upper level. However we do not know yet the transition function, $\tilde{\mathcal{E}}$, for $\tilde{\varPhi}_{T}$. Moreover, like in \cite{Israeli} the interest here is to construct $\tilde{\mathcal{E}}$ from the transition function in the lower level.  With this goal in mind, an analogous procedure  \cite{Israeli} for the PCA is proposed. 

 The first step is to apply the transition function in the lower level $h$ times, i.e.,
\begin{equation}
\label{eq:Time_rescaling}
\mathcal{E}^{h}\varPhi_{t}^{s}=\varPhi_{t+h}^{s},\ \text{with}\; h\leq s.
\end{equation}
Different from the related result by Israeli and Goldenfeld, which requires $h=s$. Here we can relax this constraint, leading to the cases we denote by \emph{temporal and spatial coarse graining}. Subsequently, the CG map is applied to get a PCA state in the ``upper level'' at time $T+1$, that is,
\begin{equation}
\label{eq:proj_t+h}
\varLambda_{CG}^{\left|L\right|/s}\varPhi_{t+h}^{s}=\tilde{\varPhi}_{T+1}.
\end{equation}
Then, we say that a PCA in the upper level is emergent from the lower level, as long as there exists a PCA transition function $\tilde{\mathcal{E}}$ satisfying the PCA definition for transition functions, i.e, composed by local operators, that connects these two PCA states. Mathematically speaking, we are looking for a transition function
\begin{equation}
\label{eq:upperTrans}
\tilde{\mathcal{E}}(\tilde{\varPhi}_{T})=\tilde{\varPhi}_{T+1}.
\end{equation}
Besides being a valid PCA transition function, it must also observe that given any two  distinct states  $\varPhi_{t}^{s}$ and $\Theta_{t}^{s}$, such that $\varLambda_{CG}^{\left|L\right|/s}\left(\varPhi_{t}^{s}\right)=\varLambda_{CG}^{\left|L\right|}/s\left(\Theta_{t}^{s}\right)$, then
\begin{eqnarray}
\label{eq:constraint_pca}
\tilde{\mathcal{E}}\left(\varLambda_{CG}^{\left|L\right|/s}\varPhi_{t}^{s}\right)&=&\varLambda_{CG}^{\left|L\right|/s}\left(\mathcal{E}^{h}\varPhi_{t}^{s}\right)\\
&=&\varLambda_{CG}^{\left|L\right|/s}\left(\mathcal{E}^{h}\Theta_{t}^{s}\right)=\tilde{\mathcal{E}}\left(\varLambda_{CG}^{\left|L\right|/s}\Theta_{t}^{s}\right).\nonumber
\end{eqnarray}

So far we have described the CG procedure acting in the PCA state, that includes all supercells Eq.(\ref{eq:SperCellPCA}). However, from the PCA space homogeneity and from its time and space translation invariance, the procedure can be done just by analyzing the states within the neighborhood scheme.

Notice that $h>s$ is not allowed, since in these cases there will be enough time for the excitation to cross the neighborhood scheme in the upper level. This restriction can be better understood with a simple example. Let us choose $s=2$ with the following neighborhood scheme $\mathcal{N}_{x}=\left\{ x-1,x,x+1\right\}$. Now, if we were to choose  $h>2$, e.g., $h=3$ the excitation can arrive in the supercell $x\pm2$, and thus out the neighborhood scheme. In this case the procedure will fail, since the transition function in the upper level only interacts inside $\mathcal{N}=\left\{\tilde{x}-1,\tilde{x},\tilde{x}+1\right\}$ where $\tilde{x}$ refers to the position in the upper level. Then, by allowing $h>s$, there is a chance of an emergent structure with non-local operators to appear, i.e., a transition function that interacts the cells $\tilde{x}$ and $\tilde{x}\pm2$ not respecting the neighborhood scheme in the upper level. 

\begin{figure}[ht]
	\includegraphics[trim=100 80 60 60,clip,scale=0.65]{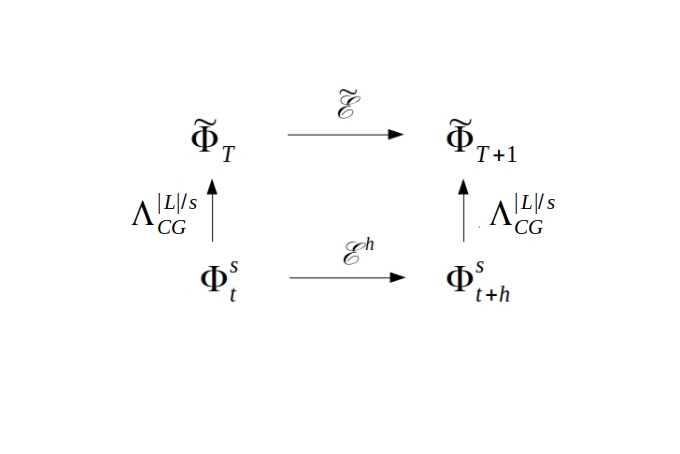}\caption{\small\label{fig:PCA_scheme}Schematic diagram summarizing the general procedure.}
\end{figure}

The general procedure can be summarized in the scheme presented in Fig.~\ref{fig:PCA_scheme}. 

At this point it is important to discuss some characteristics of the $\Lambda_{CG}$ employed in this model.  From Eq.(\ref{eq:first_part}) and from the fact that the same number of subcells are kept in both levels (i.e., $n'=n$), it turns out that $\Lambda_{CG}$ is a map $\Lambda_{CG}:\mathbb{Z}^{sn}\rightarrow \mathbb{Z}^{n}$. This means that in the deterministic cases $\Lambda_{CG}$ belongs to the space of $n\times sn$ matrices, only with 0s or 1s entries. This implies that the map is not injective, thus different states in the lower level may give the same state on the upper level. Physically speaking, there are different microscopic states that correspond to the same macroscopic state. Moreover, there is another important characteristic of the map which is a consequence of the physical interpretation we are using in our investigations. Since the interpretation used here is that the value 1 in the subcells is equivalent to the existence of one particle (or excitation), and 0 that a given location is empty,in order to preserve the number of particles during the evolution, we only allow one single value different of zero in each column and each row of $\Lambda_{CG}$. If that was not the case, the maps could increase the number of particles after coarse graining, as we illustrate in \ref{secIII}, and it could also lead to dynamics in the upper level that do not conserve the number of particles. Another relevant information for further analysis is the number of possible CG maps, $N_{CG}(n,s)$, given the supercell size $s$ and the number of subcells $n$. There are three main points that should be considered to account for the total number of maps: The size of the matrix, $n\times ns$; the constraint about the number of non-zeros;  and the fact that maps with rows and columns only with zeros are allowed. When these main informations about the problem are combined, the following number is be established

\begin{equation}
\label{eq:number_maps}
N_{CG}(n,s)=\sum_{i=0}^{n-1}\frac{ns!}{(ns-n+i)!}\binom{n}{i}.
\end{equation}
The restriction about the number of 1s in each column can be removed  when  we are restricted to a single particle scenario, which is exactly the case that we work with in what follows. By dropping this restriction the number of possible CG maps changes to 
\begin{equation}
\label{eq:number_maps2}
N_{CG}(n,s)=\left(n+1\right)^{ns}-1.
\end{equation}

Herein, only the results for the CG maps that take two and three cells ($s=2,3$) to one are reported. The extension for more dimensions and for different values for $s$ can be done naturally. 

The last point to be noticed when attempting to apply the CG to deterministic settings is the number of possible connections between the lower and upper levels. There are $n!$ permutation matrices for $n$ subcells. Moreover, the PCAs will be kept with the same structure in the lower and upper levels (i.e., the same neighborhood scheme and the same number of subcells). From each initial dynamics $\mathcal{E}(\text{Swap},\pi^{(i)})$, $n!$ permutation matrices give $n!$ possible connections to deterministic dynamics in the upper level $\tilde{\mathcal{E}}(\text{Swap},\pi^{(j)})$. Now taking in account the $n!$ different initial conditions in the lower level we conclude that there are  $\left(n!\right)^{2}$ possible \emph{links} between the lower and the upper levels. The word "link" was adopted to emphasize the existence of connections between the lower and upper levels. In a case with more than one CG map connecting the same rules between these two levels, they are counted only once, i.e., just one link between these rules. The number of links -- whose biggest value is $(n!)^2$ -- gives us the number of different rules connecting the lower and upper levels. This fact will be important in the analyses of the results, for a better understanding of its quantitative and qualitative aspects. 


\section{Coarse Graining results for one-dimensional PCA}\label{secIII}

Throughout this section we assume a neighborhood scheme $\mc{N}_x=\{x-1, \; x,\; x+1\}$.

\subsection{Deterministic results}

\subsubsection{Spatial coarse-graining}

In this first part we describe the results for the case where the temporal coarse graining is not applied, which means $\mathcal{E}^h$ with $h=1$.

\paragraph{Two cells, $s=2$, to one cell:}

Our starting point is $n=2$, for a case where there is a map from two cells to one. The idea here is to use this simple example to explain how our procedure works and to check the consistence of the method illustrated in Fig.~\ref{fig:PCA_scheme}.  Given the simplicity of this example, here we consider a one-dimensional lattice where the number of particles is arbitrary. In this case there are only two different permutation matrices,
\begin{equation}
\pi^{(1)}=\begin{pmatrix}1 & 0\\
0 & 1
\end{pmatrix},\; \pi^{(2)}=\begin{pmatrix}0 & 1\\
1 & 0
\end{pmatrix}.
\end{equation}

Then, working only with $\sigma_1=\text{Swap}$ as the local interaction operator, only the two deterministic transition functions $\mathcal{E}\left(\text{Swap},\pi^{(1)}\right)$ and $\mathcal{E}\left(\text{Swap},\pi^{(2)}\right)$ are possible.

Despite the fact that there are four possible links connecting the lower to the upper level, only the connection $\mathcal{E}\left(\text{Swap},\pi^{(1)}\right)$ to $\tilde{\mathcal{E}}\left(\text{Swap},\\\pi^{(1)}\right)$ is obtained, with the CG map given by 
\begin{equation}
\label{eq:first_cg}
\Lambda_{CG}=\begin{pmatrix}1 & 0 & 0 & 0\\
0 & 0 & 0 & 1
\end{pmatrix}.
\end{equation}

Now let us to check if these dynamics in the lower and upper level, alongside Eq.(\ref{eq:first_cg}), obey the constraints imposed by the CG procedure.  

Starting with the dynamics generated by $\mathcal{E}\left(\text{Swap},\pi^{(1)}\right)$, its consequence is to keep the particles confined between two neighboring cells, with a forward and backward movement from one to the other. This is represented as
{\footnotesize
\begin{eqnarray*}
\cdots\left(d^{(0)},d^{(1)}\right)_{x}\left(e^{(0)},e^{(1)}\right)_{x+1}\cdots\overset{\mathcal{E}\left(\text{Swap},\pi^{(1)}\right)}{\rightleftarrows}\\
\cdots\left(c^{(1)},e^{(0)}\right)_{x}\left(d^{(1)},f^{(0)}\right)_{x+1}\cdots,
\end{eqnarray*}}
where we assigned each lattice site by a Boolean variable with superscripts related to the subcells location at the current time step. As we can see from this dynamics above  a single particle in the right-most subcell of cell $x$ moves to the left-most subcell of cell $x+1$,  and vice-versa.

Now, let us compose a supercell by putting cells $x$ and $x+1$ together and then applying the CG map (\ref{eq:first_cg}) before the transition function
{\footnotesize
\begin{eqnarray*}
&&\Lambda_{CG}^{3}\left[\left\{ \left(b^{(0)},b^{(1)}\right)_{x-2}\left(c^{(0)},c^{(1)}\right)_{x-1}\right\} ,\right.\\
&&\left.\left\{ \left(d^{(0)},d^{(1)}\right)_{x_i}\left(e^{(0)},e^{(1)}\right)_{x+1}\right\} ,\left\{\left(f^{(0)},f^{(1)}\right)_{x+2}\left(g^{(0)},g^{(1)}\right)_{x+3}\right\} \right],
\end{eqnarray*}}
that give us
\beq
\label{eq:CG_ini}
\left(b^{(0)},c^{(1)}\right)_{\tilde{x}_i-1},\left(d^{(0)},e^{(1)}\right)_{\tilde{x}},\left(f^{(0)},g^{(1)}\right)_{\tilde{x}+1}.
\eeq
The transition function is then applied before the CG map, to the same initial state
{\footnotesize
\begin{eqnarray*}
\mathcal{E}\left[ \left\{ \left(b^{(0)},b^{(1)}\right)_{x-2}\left(c^{(0)},c^{(1)}\right)_{x-1},\right\} ,\left\{ \left(d^{0},d^{1}\right)_{x_i}\left(e^{(0)},e^{(1)}\right)_{x+1}\right\} \right.\\
\left\{ \left(f^{(0)},f^{(1)}\right)_{x}\left(g^{(0)},g^{(1)}\right)_{x+1}\right\}\\
=\left[ \left\{ \left(a^{(1)},c^{0}\right)_{x-2}\left(b^{(1)},d^{(0)}\right)_{x-1},\right\} ,\left\{ \left(c^{(1)},e^{(0)}\right)_{x_i}\left(d^{(1)},f^{(0)}\right)_{x+1}\right\} \right.\\
\left\{ \left(e^{(1)},g^{(0)}\right)_{x}\left(f^{(1)},h^{(0)}\right)_{x+1}\right\},
\end{eqnarray*}}
and after the CG map,
\beq
\label{eq:CG_evo}
\left(a^{(1)},d^{(0)}\right)_{\tilde{x}-1},\left(c^{(1)},f^{(0)}\right)_{\tilde{x}},\left(e^{(1)},h^{(0)}\right)_{\tilde{x}+1}.
\eeq
Therefore, we can see that the results \ref{eq:CG_ini} and \ref{eq:CG_evo} are compatible with the upper transition rule $\tilde{\mathcal{E}}\left(\text{Swap},\pi^{(1)}\right)$. Furthermore, within this simple example, by changing our CG map in Eq.(\ref{eq:CG_ini}) to
\begin{equation}
\label{eq:Exa_cg}
\Lambda_{CG}=\begin{pmatrix}1 & 1 & 0 & 0\\
0 & 0 & 0 & 1
\end{pmatrix},
\end{equation}	 
the argument for only one single value different from zero at each row can be verified. Rather then Eq.(\ref{eq:CG_ini}) we would get
{\footnotesize
\[
\left(b^{(0)}+b^{(1)},c^{(1)}\right)_{\tilde{x}-1},\left(d^{(0)}+d^{(1)},e^{(1)}\right)_{\tilde{x}},\left(f^{(0)}+f^{(1)},g^{(1)}\right)_{\tilde{x}+1},
\]}
that would create subcells with two particles, that we do not allow here. However, we can also notice that the map given in Eq.(\ref{eq:Exa_cg}) is only problematic when we are in the scenario of multiple particles. Within the same example, by changing our CG map now to 
\begin{equation}
\label{eq:Exa_cg2}
\Lambda_{CG}=\begin{pmatrix}1 & 0 & 0 & 0\\
1 & 0 & 0 & 1
\end{pmatrix},
\end{equation}	
we can also verify the constraint in the rows should be respected. Now \st{ rather that} we would get
{\footnotesize
\[
\left(b^{(0)},b^{(0)}+c^{(1)}\right)_{\tilde{x}-1},\left(d^{(0)},d^{(0)}+e^{(1)}\right)_{\tilde{x}},\left(f^{(0)},f^{(0)}+g^{(1)}\right)_{\tilde{x}+1}.
\]}
In this scenario, even starting with a single  excitation, for instance by setting up all bits to zero except $b^{(0)}=1$, the problem would remain. At the end of the process we would get two particles.

Moving on to the computational task to get these results, there are two possibilities to deal with this problem. One is to fix the transition function in the lower and the upper levels, and then search for a CG map that would connect both levels. The second option is to fix both the transition function in the lower level and a CG map and then to look for a transition function in the upper level. Let us show one example using this last procedure.  In order to make this illustration simpler all the subcells values are set to zero, except $d^{(1)}$ which is set to one.  Beginning with the state $\varPhi^2_t\in\mathbb{Z}_{2}^{4}\times\mathbb{Z}_{2}^{4}\times\mathbb{Z}_{2}^{4}$, with three supercells in the case for $n=2$,
\[
\varPhi_{t}^{2}=\begin{pmatrix}0_{4}\\
1\\
0\\
0\\
0\\
0_{4}
\end{pmatrix},
\] 
where the subscript 4 refers to the vector composed of four 0s
\[
0_{4}=\begin{pmatrix}0\\
0\\
0\\
0
\end{pmatrix}.
\] 
As the next step, the CG map in Eq.(\ref{eq:first_cg}), which is a $2\times4$ matrix, is applied
\[
\begin{pmatrix}\Lambda_{CG}\\
& \Lambda_{CG}\\
&  & \Lambda_{CG}
\end{pmatrix}\begin{pmatrix}0_{4}\\
1\\
0\\
0\\
0\\
0_{4}
\end{pmatrix}=\begin{pmatrix}0\\
0\\
1\\
0\\
0\\
0
\end{pmatrix}=\tilde{\varPhi}_{T},
\]
where $\tilde{\varPhi}_{T}$ corresponds to the state with only three cells after the coarse graining  $\varPhi_t$. All matrices that act on states in the upper and lower levels are block diagonal. Returning to $\varPhi_t^2$ and applying the transition function, 

{\footnotesize
	\[
	\begin{pmatrix}1\\
	& \text{Swap}\\
	&  & \text{Swap}\\
	&  &  & \text{Swap}\\
	&  &  &  & \text{Swap}\\
	&  &  &  &  & \text{Swap}\\
	&  &  &  &  &  & 1
	\end{pmatrix}
	\]
	\[\times\begin{pmatrix}\pi^{\left(1\right)}\\
	& \pi^{\left(1\right)}\\
	&  & \pi^{\left(1\right)}\\
	&  &  & \pi^{\left(1\right)}\\
	&  &  &  & \pi^{\left(1\right)}\\
	&  &  &  &  & \pi^{\left(1\right)}
	\end{pmatrix}\begin{pmatrix}0_{4}\\
	1\\
	0\\
	0\\
	0\\
	0_{4}
	\end{pmatrix}=\begin{pmatrix}0\\
	0\\
	0\\
	1\\
	0_{4}\\
	0_{4}
	\end{pmatrix}=\varPhi_{t+1}^2,
	\]
}
with the 1s in the matrix that contains the Swap operator meaning that no operator is being applied in the boundaries, respecting the neighbor scheme. Subsequently, the CG map is applied to the state $\varPhi_{t+1}^2$,
\[
\begin{pmatrix}\Lambda_{CG}\\
& \Lambda_{CG}\\
&  & \Lambda_{CG}
\end{pmatrix}\begin{pmatrix}0\\
0\\
0\\
1\\
0_{4}\\
0_{4}
\end{pmatrix}=\begin{pmatrix}0\\
1\\
0\\
0\\
0\\
0
\end{pmatrix}=\tilde{\varPhi}_{T+1}.
\]
From ${\varPhi}_{T}$ and ${\varPhi}_{T+1}$ we can start looking whether there is some transition function in the upper level. Since the target is a permutation operator in the upper level that connects these two states, the parameterization of this operator can be done as follows,
\[
\pi^{(x)}=\begin{pmatrix}p & q\\
q & p
\end{pmatrix},
\]
where $x=1(2)$ if $p=1(0)$ and $q=0(1)$. The transition function using $\pi^{(x)}$ can now be applied, building the linear system
\[
\begin{pmatrix}1\\
& \text{Swap}\\
&  & \text{Swap}\\
&  &  & 1
\end{pmatrix}\begin{pmatrix}\pi^{\left(x\right)}\\
& \pi^{\left(x\right)}\\
&  & \pi^{\left(x\right)}
\end{pmatrix}\begin{pmatrix}0\\
0\\
1\\
0\\
0\\
0
\end{pmatrix}=\begin{pmatrix}0\\
1\\
0\\
0\\
0\\
0
\end{pmatrix}.
\]
From this simple case, $p=1$ and $q=0$ is achieved. However, this does not guarantee that there is a PCA in the upper level. In order to check that, the same procedure is performed taking into account all different states inside the supercell. Only when $p=1$ and $q=0$ are established for all distinct initial conditions is that we can say there is a PCA in the upper level, with $\tilde{\mathcal{E}}(\text{Swap},\pi^{(1)})$ as its transition function.

One of the main characteristics of our results is that the CG maps do not necessarily preserve the number of particles, what only can be noticed for some initial conditions. For instance, applying the map Eq.(\ref{eq:first_cg}) either to $\left\{ \left(0,0\right)_{x_i},\left(1,0\right)_{x_i+1}\right\}  $ or $\left\{ \left(0,1\right)_{x_i},\left(0,0\right)_{x_i+1}\right\} $, we get $\left(0,0\right)_{\tilde{x}_i}$. This is a consequence of the mathematical structure of the CG maps which are not bijective and are always reducing the number of cells. Because of that, some information loss might be expected, and here this loss is represented by the number of particles.

Moving on to the case with $n=3$, there are six different permutation matrices, as listed below:

\begin{eqnarray}
\label{eq:permutation}
\pi^{(1)}=\begin{pmatrix}1 & 0 & 0\\
0 & 1 & 0\\
0 & 0 & 1
\end{pmatrix};\:&&\pi^{(2)}=\begin{pmatrix}1 & 0 & 0\\
0 & 0 & 1\\
0 & 1 & 0
\end{pmatrix};\\\pi^{(3)}=\begin{pmatrix}0 & 1 & 0\\
1 & 0 & 0\\
0 & 0 & 1
\end{pmatrix};\:&&\pi^{(4)}=\begin{pmatrix}0 & 1 & 0\\
0 & 0 & 1\\
1 & 0 & 0
\end{pmatrix}\nonumber;\\\pi^{(5)}=\begin{pmatrix}0 & 0 & 1\\
1 & 0 & 0\\
0 & 1 & 0
\end{pmatrix};\:&&\pi^{(6)}=\begin{pmatrix}0 & 0 & 1\\
0 & 1 & 0\\
1 & 0 & 0
\end{pmatrix}.\nonumber
\end{eqnarray}
In this scenario, 12 connections are achieved from all the possible 36. These links were made by only eight distinct CG maps that are listed in Appendix \ref{h=1}.
These results are summarized in Fig. \ref{fig:resul_s2_n3}.

\begin{figure}[ht]
	\noindent \includegraphics[scale=0.45]{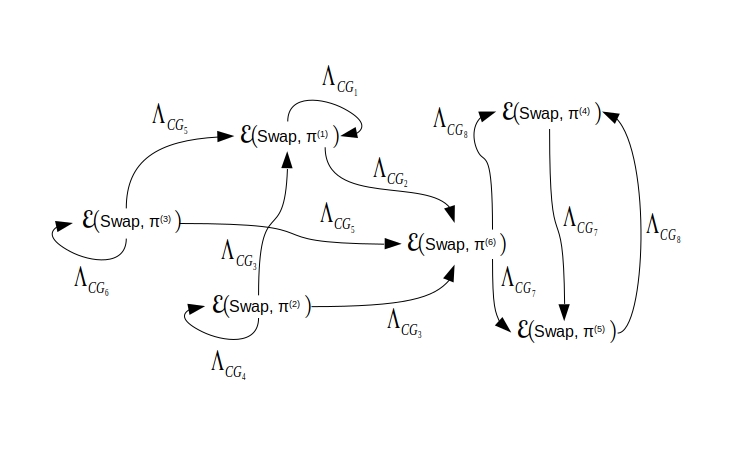}\caption{\small\label{fig:resul_s2_n3}CG results with $s=2$ and $n=3$. In this illustration we have all possible transitions functions when there are three subcells. These arrows are connecting the CA dynamics after the CG maps. For instance, beginning in the lower level with the dynamics driven by  $\pi^{(2)}$ as the permutation operator it is possible to achieve two others dynamics in the upper level, namely $\pi^{(1)}$ and $\pi^{(6)}$. Moreover, there is a map that allows the same dynamics $\pi^{(2)}$ to emerge in the upper level.}
\end{figure}

\begin{figure}[ht]
	\noindent \centering\includegraphics[scale=0.5]{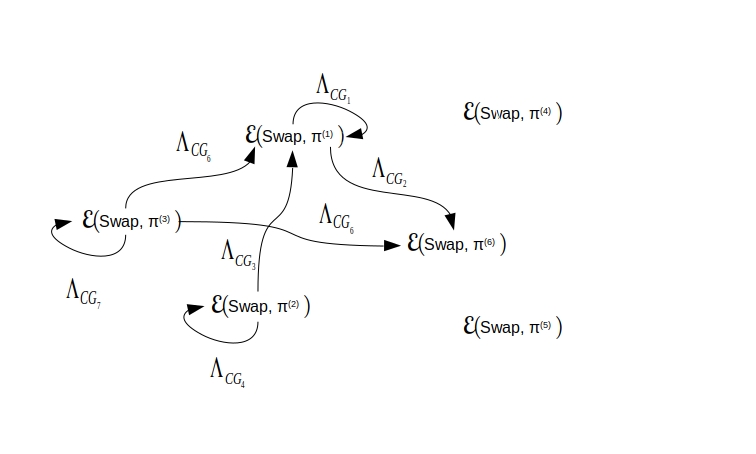}\caption{\small\label{fig:resul_s3_h3}CG results with $s=3$ and $n=3$.}
\end{figure}

\paragraph{Three cells, $s=3$, to one cell:} We again start with $n=2$. Our results in this case are just an extension of the previous one, since their transition functions in the lower and the upper levels yield the same dynamics. One link between the same transition functions $\mathcal{E}\left(\text{Swap},\pi^{(1)}\right)$ to $\tilde{\mathcal{E}}\left(\text{Swap},\pi^{(1)}\right)$ is established, and the CG map that establishes this link is equivalent to Eq.(\ref{eq:first_cg}) in this new space dimension, namely 
\begin{equation}
\label{first_cg}
\Lambda_{CG}=\begin{pmatrix}1 & 0 & 0 & 0 & 0 & 0\\
0 & 0 & 0 & 0 & 0 & 1
\end{pmatrix}.
\end{equation}
Now with $n=3$, i.e three subcells, 8 links are obtained, out of 36 possibilities. These connections are given by seven CG maps (see Section \ref{s=3,h=1}).
The results are illustrated in Fig. \ref{fig:resul_s3_h3}.


\subsubsection{Spatial and temporal coarse-graining}
The purely spatial setting was defined as the cases where $h=1$. Here we open to more possibilities for the number of times which the transition function can be applied, respecting the bound $h\leq s$ that we showed before. Since the time step in the lower level is given by the $h$ value the immediate consequence of doing that is the time flowing in these levels is different and because of that we call these cases by \textit{temporal coarse-graining}.

\paragraph{Two cells, $s=2$, to one cell:}
From the previous bound in this case we can use only $h=2$, since the case with $h=1$ has already been explored. 

\begin{figure}[ht]
	\noindent \centering\includegraphics[scale=0.5]{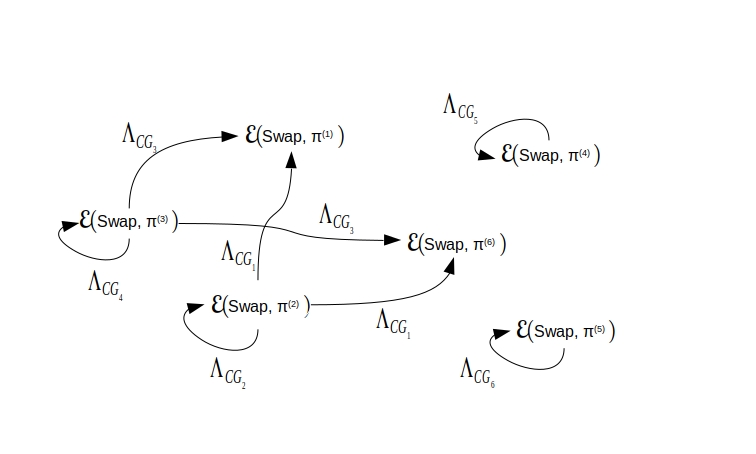}\caption{\small\label{fig:h2resul_s2_n3} CG results with $s=h=2$ and $n=3$.}
\end{figure}

Starting with $n=2$, no link is possible between the lower and upper levels. But with $n=3$, 8 links are possible, out of six different maps:
\begin{eqnarray*}
	\Lambda_{CG_{1}}=\begin{pmatrix}0 & 0 & 0 & 0 & 0 & 0\\
		0 & 1 & 1 & 1 & 0 & 0\\
		0 & 0 & 0 & 0 & 0 & 0
	\end{pmatrix};\:&&\Lambda_{CG_{2}}=\begin{pmatrix}1 & 0 & 0 & 0 & 0 & 0\\
		0 & 0 & 0 & 0 & 0 & 1\\
		0 & 0 & 0 & 0 & 1 & 0
	\end{pmatrix};\\\Lambda_{CG_{3}}=\begin{pmatrix}0 & 0 & 0 & 0 & 0 & 0\\
		0 & 0 & 1 & 1 & 1 & 0\\
		0 & 0 & 0 & 0 & 0 & 0
	\end{pmatrix};\:&&\Lambda_{CG_{4}}=\begin{pmatrix}0 & 1 & 0 & 0 & 0 & 0\\
		1 & 0 & 0 & 0 & 0 & 0\\
		0 & 0 & 0 & 0 & 0 & 1
	\end{pmatrix};\\\Lambda_{CG_{5}}=\begin{pmatrix}1 & 0 & 0 & 0 & 0 & 0\\
		0 & 1 & 1 & 0 & 0 & 0\\
		0 & 0 & 0 & 0 & 1 & 1
	\end{pmatrix};\:&&\Lambda_{CG_{6}}=\begin{pmatrix}1 & 1 & 0 & 0 & 0 & 0\\
		0 & 0 & 0 & 0 & 0 & 0\\
		0 & 0 & 0 & 0 & 0 & 1
	\end{pmatrix}.
\end{eqnarray*}
These are fewer connections and maps than previously established with $h=1$. These results are summed up in Fig. \ref{fig:h2resul_s2_n3}.

\begin{figure}[ht]
	\noindent \centering\includegraphics[scale=0.5]{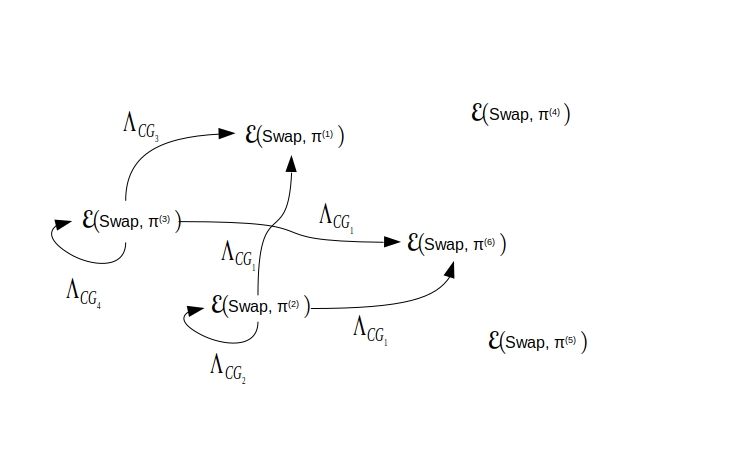}\caption{\small\label{fig:h2resul_s3_n3}CG results for $s=3$ with $h=2$ and $n=3$.}
\end{figure}

\paragraph{Three cells, $s=3$, to one cell:}
Now we can work with the two values $h=2$ and $h=3$, since $s=3$. Beginning with $h=2$, no result is obtained with $n=2$; with $n=3$ the results are summarized in Fig. \ref{fig:h2resul_s3_n3}.
In this case we only got four different maps \ref{s=3,h=2}.
\begin{figure}[ht]
	\noindent \centering\includegraphics[scale=0.5]{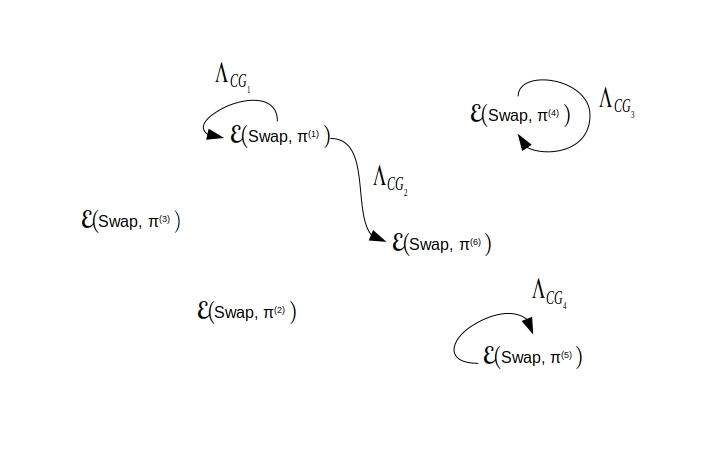}\caption{\small\label{fig:h3resul_s3_n3} CG results for $s=h=3$ and $n=3$.}
\end{figure}
Moving on to $h=3$, with $n=2$ the result is the same as that with $h=1$, since $\mathcal{E}^1=\mathcal{E}^3$. With $n=3$, again four maps are achieved \ref{s=3,h=3} but now only four links, one map to each link Fig. \ref{fig:h3resul_s3_n3}. 

\subsubsection{Overview of deterministic results}

\begin{figure}[ht]
	\noindent \centering\includegraphics[scale=0.55]{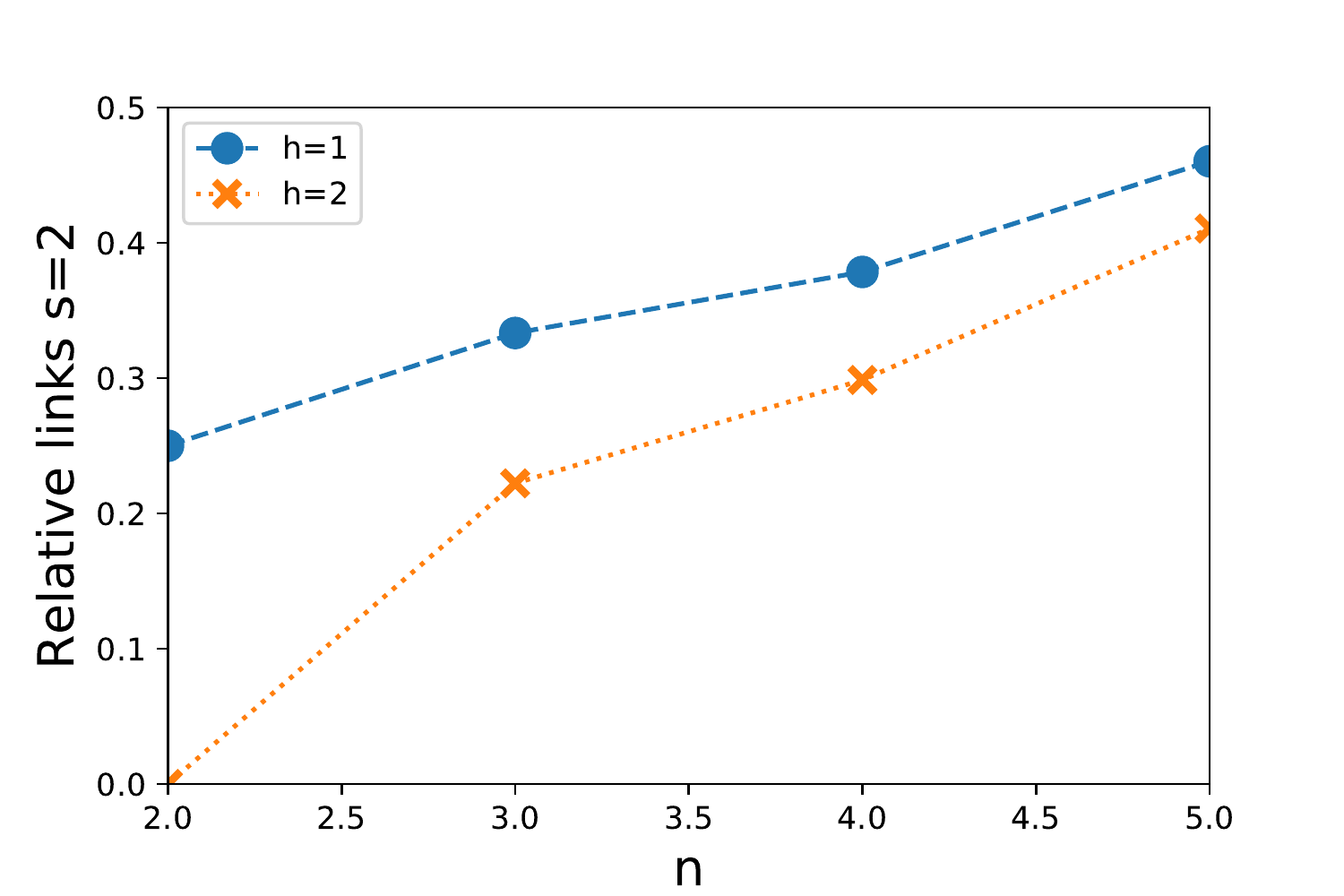}\caption{\small\label{fig:s2}Relative links from two cells to one cell, after application of the CG map. These results point to us that more and more dynamics become accessible as we increase the number of subcells in both scenarios: spatial and temporal coarse-graining.}
\end{figure}

\begin{figure}[ht]
	\noindent \centering\includegraphics[scale=0.55]{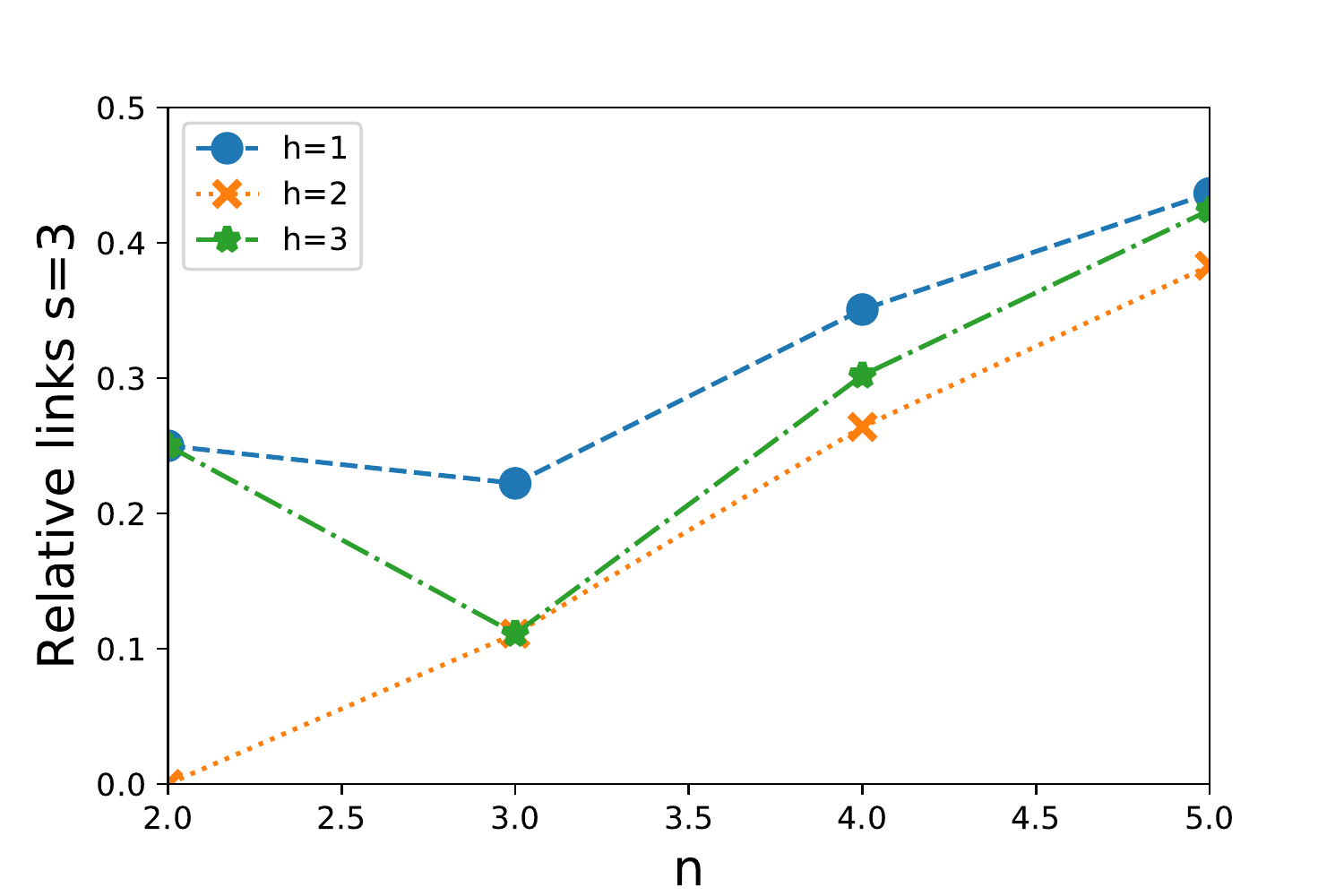}\caption{\small\label{fig:s3}Relative links from three cells to one cell, after the application of the CG map.}
\end{figure}

During our investigations, the cases with $n=4$ and $n=5$ were considered in order to observe the consequence of the \textit{relative links}, i.e. the links achieved over the total of possible links, as Figs. \ref{fig:s2} and \ref{fig:s3} display. From all the results on deterministic dynamics in the lower level it is not possible to have all dynamics in the upper level as emergent ones. However, the number of relative links increases, as shown in Fig. \ref{fig:s2} for $s=2$, and in Fig. \ref{fig:s3} for $s=3$.
Should we expect all links to appear for some number of subcells? In fact, this seems quite likely. Let us understand why by considering either Eq.(\ref{eq:number_maps}) or  Eq.(\ref{eq:number_maps2}), the total number of possible maps given by $n$ and $s$. Since $\left(n!\right)^{2}$ is the number of possible links, this means that the number of maps increases faster than the total number of links, for a given $s$, i.e.,
\[
\lim_{n\rightarrow\infty}\frac{\left(n!\right)^{2}}{N_{CG}(n,s)}=0.
\]
Consequently, we can expect that, at some point, all links will appear. Since the number of subcells are increasing, the number of maps increases faster than the possible dynamics. In the end, it means that the space dimension, i.e., the number of cells, is contracting more and more, ultimately implying that the existence of more microscopic dynamics that cannot be distinguished after coarse graining.

Another observation is that the number of links also depends on the values of $h$ employed. The reason is that there are values for $h$ -- the number of times that the transition function should be applied in the lower level before the state is coarse grained -- that might lead the particle to stay inside the same initial supercell. In these cases, the trivial dynamics is established in the upper level, which means particles that do not move to their neighbors. Once these possibilities are not included, fewer links become available for these cases.


\subsection{Stochastic CG results for one-dimensional PCA}

Until now a strict constraint was made in the dynamics after the coarse-graining. The imposition was such that only permutation operators are allowed in the upper level. However, it might be the case that these constraints are too artificial to real physical systems, which can explain why it is so difficult to find CG maps linking two deterministic dynamics. 

\begin{figure}[ht]
	\noindent \includegraphics[trim=80 80 60 60,clip,scale=0.5]{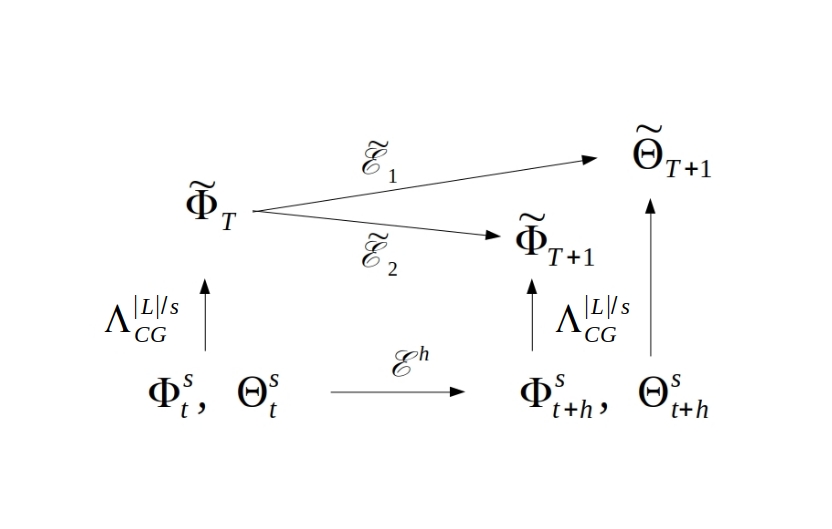}\caption{\small\label{fig:PCA_scheme2}In this illustration, we have two different states in the lower level, namely  $\Phi_t^s$ and $\Theta_t^s$ that represents the same state in the upper level $\tilde{\Theta}_T$. Then, these two states in the lower level evolve to two different states as well. So far there is nothing new in this procedure compared with the construction did to deterministic case. However,  without the constraint given in Eq.(\ref{eq:constraint_pca}) it is allowed that these two states at time $t+h$ go to two different states in the upper level. Because of that, different transition functions in the upper level can appear.  }
\end{figure}

Since it is quite common in physics to deal with stochastic dynamics when we do not have access to the full information about the system, it seems more genuine to search for convex combination of permutations in the upper level, starting from some fixed CA dynamics in the lower level, which is our next step. This is possible as long as the specific constraint previously on $\tilde{\mathcal{E}}$, according to Eq.(\ref{eq:constraint_pca}) imposed to achieve the transition function, is relaxed. Without that constraint, if we have two or more initial states in the lower level leading to the same state after the coarse graining they might evolve to different states in the upper level. Thus, at the end of this process, different transition functions in the upper level are possible. This idea can be visualized in Fig. \ref{fig:PCA_scheme2}.

Alternatively of what we saw previously, here there is the possibility  to get different maps from the same link. Thus, it allows for more maps in comparison with the previous results. In what follows, we only give the results for $s=2$ with three subcells; also, since the permutation matrices are the same ones employed in the deterministic cases, the notation used in Eq.\ref{eq:permutation} is kept.

Here we call the attention to the strong physical meaning achieved to the stochastic cases, where the list of results can be checked in Appendix \ref{stochastic_Results}.  We can point out the general structure of the transition function achieved in the upper level. For the first tiling, we often got
\begin{equation}
\label{eq:Brownian}
\sigma_0=p_{1}\pi^{(1)}+p_{6}\pi^{(6)},
\end{equation}  
where $p_{1}, p_{6}\geq0$ and $p_{1}+p_{6}=1$, and the swap operator for the second tiling. The interesting point about this dynamics is the fact that it describes the Random Walk (RW) problem~\cite{RW}. To see that let us recall how this problem is described.

In the RW problem at every point, before its displacement at the one-dimensional lattice, the walker flips a coin. In this case, the coin is related to the walker probability to keep moving to the same direction or change its movement direction. Thus, here we can see Eq.(\ref{eq:Brownian}) playing the coin's role, ($p_1$ is the walker probability to not change its direction and $p_6$ the probability to change) and the shift operator giving the walker displacement, agreeing with what we claimed above.

Therefore, this discrete equation of motion is the one that gives a stochastic partial differential equation
\[
\partial_{t}\rho+D\partial_{x}^{2}\rho=0,
\]
in the continuous limit \cite{cellubook}, where $\rho(x,t)$ is the local density of particles,  $D$ is the
diffusion constant given by 
\begin{equation}
D=\frac{\lambda^{2}}{\tau}\left(\frac{p_{1}}{2\left(1-p_{1}\right)}\right),
\end{equation}
and $\lambda^2/\tau$ is a constant that comes from the dispersion relation of the problem.


\section{Conclusion}\label{secConc}
Similarly to \cite{Israeli}, in the present work we studied emergent dynamics, but in a different scenario of CA. Differently from the previous results, with PCA we could get CG maps in different time scales. One advantage of this CA class is its strong connection with physical processes; for instance, the Navier Stokes equation \cite{FHP} and Random Walk \cite{cellubook} can be simulated by applying this computation model. Moreover, we established two distinct results: links connecting deterministic CA to deterministic CA, and deterministic CA to stochastic CA.

Despite the fact that the results in the deterministic cases suggest that all links between the lower and the upper levels will be achieved, for some large number of subcells, we could see how difficult it is to get these emergent phenomena, since the total number of CG maps increases much faster than the number of possible links, as the number of subcells increases. Another point that should be considered is that, while we could not observe different CG maps linking two different transition functions in the deterministic results, it happened very often in the stochastic cases. This can be interpreted as an indication to why stochastic processes in the macroscopic world naturally emerge from well determined individual particle actions, in agreement with statistical mechanics.

By taking advantage of the PCA, the last section showed that the procedure introduced here can be easily translated to the case of multiple particles, sufficing to be more careful when there is interaction between them. 

Going beyond the classical CA, the CG prescription might be translated to its quantum counterpart. Instead of the CA explored in \cite{Wolfram,Israeli}, it is the PCA that should be quantized to get the partitioned unitary quantum cellular automata (PUQCA,  \cite{PUQCA}). A core part that does the PCA be its classical analogous is the reversibility established when we have only permutation operators acting, since the unitary evolution in QCA makes it reversible at any time. Therefore, rather than extending the method showed in \cite{Israeli} to the QCA, a natural choice is picking up the prescription introduced here and extend it to its quantum version. In quantum theory this tool can be useful, for instance, to study the transition from the quantum to the classical world~\cite{duarte2017emerging,zurekRMP}.

\section*{Acknowledgements}
\noindent We acknowledge financial support from the National Institute for Science and Technology of Quantum Information (INCT-IQ/CNPq) and CAPES, both from Brazil. We also would like to thank Pedro De Oliveira for careful reading of the manuscript.

\section{Appendices}

\subsection{CG maps for $s=2$ and $n=3$}

\subsubsection{h=1}\label{h=1}

\begin{eqnarray*}
	\Lambda_{CG_{1}}=\begin{pmatrix}1 & 0 & 0 & 0 & 0 & 0\\
		0 & 0 & 0 & 0 & 0 & 0\\
		0 & 0 & 0 & 0 & 0 & 1
	\end{pmatrix};\:&&\Lambda_{CG_{2}}=\begin{pmatrix}0 & 0 & 0 & 0 & 0 & 0\\
		0 & 1 & 1 & 1 & 1 & 0\\
		0 & 0 & 0 & 0 & 0 & 0
	\end{pmatrix};\\\Lambda_{CG_{3}}=\begin{pmatrix}0 & 0 & 0 & 0 & 0 & 0\\
		0 & 1 & 1 & 1 & 0 & 0\\
		0 & 0 & 0 & 0 & 0 & 0
	\end{pmatrix};\:&&\Lambda_{CG_{4}}=\begin{pmatrix}1 & 0 & 0 & 0 & 0 & 0\\
		0 & 0 & 0 & 0 & 1 & 0\\
		0 & 0 & 0 & 0 & 0 & 1
	\end{pmatrix};\\\Lambda_{CG_{5}}=\begin{pmatrix}0 & 0 & 0 & 0 & 0 & 0\\
		0 & 0 & 1 & 1 & 1 & 0\\
		0 & 0 & 0 & 0 & 0 & 0
	\end{pmatrix};\:&&\Lambda_{CG_{6}}=\begin{pmatrix}1 & 0 & 0 & 0 & 0 & 0\\
		0 & 1 & 0 & 0 & 0 & 0\\
		0 & 0 & 0 & 0 & 0 & 1
	\end{pmatrix};\\\Lambda_{CG_{7}}=\begin{pmatrix}0 & 0 & 0 & 0 & 0 & 0\\
		0 & 0 & 1 & 0 & 0 & 0\\
		0 & 0 & 0 & 0 & 0 & 1
	\end{pmatrix};\:&&\Lambda_{CG_{8}}=\begin{pmatrix}1 & 0 & 0 & 0 & 0 & 0\\
		0 & 0 & 0 & 1 & 0 & 0\\
		0 & 0 & 0 & 0 & 0 & 0
	\end{pmatrix}.
\end{eqnarray*}

\subsubsection{h=2}\label{h=2}

\begin{eqnarray*}
	\Lambda_{CG_{1}}=\begin{pmatrix}0 & 0 & 0 & 0 & 0 & 0\\
		0 & 1 & 1 & 1 & 0 & 0\\
		0 & 0 & 0 & 0 & 0 & 0
	\end{pmatrix};\:&&\Lambda_{CG_{2}}=\begin{pmatrix}1 & 0 & 0 & 0 & 0 & 0\\
		0 & 0 & 0 & 0 & 0 & 1\\
		0 & 0 & 0 & 0 & 1 & 0
	\end{pmatrix};\\\Lambda_{CG_{3}}=\begin{pmatrix}0 & 0 & 0 & 0 & 0 & 0\\
		0 & 0 & 1 & 1 & 1 & 0\\
		0 & 0 & 0 & 0 & 0 & 0
	\end{pmatrix};\:&&\Lambda_{CG_{4}}=\begin{pmatrix}0 & 1 & 0 & 0 & 0 & 0\\
		1 & 0 & 0 & 0 & 0 & 0\\
		0 & 0 & 0 & 0 & 0 & 1
	\end{pmatrix};\\\Lambda_{CG_{5}}=\begin{pmatrix}1 & 0 & 0 & 0 & 0 & 0\\
		0 & 1 & 1 & 0 & 0 & 0\\
		0 & 0 & 0 & 0 & 1 & 1
	\end{pmatrix};\:&&\Lambda_{CG_{6}}=\begin{pmatrix}1 & 1 & 0 & 0 & 0 & 0\\
		0 & 0 & 0 & 0 & 0 & 0\\
		0 & 0 & 0 & 0 & 0 & 1
	\end{pmatrix}.
\end{eqnarray*}

\subsection{CG maps for $s=3$ and $n=3$}

\subsubsection{h=1}\label{s=3,h=1}

{\footnotesize
	\begin{eqnarray*}
		\Lambda_{CG_{1}}=\begin{pmatrix}1 & 0 & 0 & 0 & 0 & 0 & 0 & 0 & 0\\
			0 & 1 & 1 & 1 & 1 & 1 & 1 & 1 & 0\\
			0 & 0 & 0 & 0 & 0 & 0 & 0 & 0 & 1
		\end{pmatrix};\:&&\Lambda_{CG_{2}}=\begin{pmatrix}0 & 0 & 0 & 0 & 0 & 0 & 0 & 0 & 0\\
			0 & 1 & 1 & 1 & 1 & 1 & 1 & 1 & 0\\
			0 & 0 & 0 & 0 & 0 & 0 & 0 & 0 & 0
		\end{pmatrix};\\\Lambda_{CG_{3}}=\begin{pmatrix}0 & 0 & 0 & 0 & 0 & 0 & 0 & 0 & 0\\
			0 & 1 & 1 & 1 & 1 & 1 & 1 & 0 & 0\\
			0 & 0 & 0 & 0 & 0 & 0 & 0 & 0 & 0
		\end{pmatrix};\:&&\Lambda_{CG_{4}}=\begin{pmatrix}1 & 0 & 0 & 0 & 0 & 0 & 0 & 0 & 0\\
			0 & 0 & 0 & 0 & 0 & 0 & 0 & 1 & 0\\
			0 & 0 & 0 & 0 & 0 & 0 & 0 & 0 & 1
		\end{pmatrix};\\\Lambda_{CG_{5}}=\begin{pmatrix}0 & 0 & 0 & 0 & 0 & 0 & 0 & 0 & 0\\
			0 & 1 & 1 & 1 & 1 & 1 & 1 & 0 & 0\\
			0 & 0 & 0 & 0 & 0 & 0 & 0 & 0 & 0
		\end{pmatrix};\:&&\Lambda_{CG_{6}}=\begin{pmatrix}0 & 0 & 0 & 0 & 0 & 0 & 0 & 0 & 0\\
			0 & 0 & 1 & 1 & 1 & 1 & 1 & 1 & 0\\
			0 & 0 & 0 & 0 & 0 & 0 & 0 & 0 & 0
		\end{pmatrix};\\\Lambda_{CG_{7}}=\begin{pmatrix}1 & 0 & 0 & 0 & 0 & 0 & 0 & 0 & 0\\
			0 & 1 & 0 & 0 & 0 & 0 & 0 & 0 & 0\\
			0 & 0 & 0 & 0 & 0 & 0 & 0 & 0 & 1
		\end{pmatrix}.&&
\end{eqnarray*}}

\subsubsection{h=2}\label{s=3,h=2}

{\footnotesize
	\begin{eqnarray*}
		\Lambda_{CG_{1}}=\begin{pmatrix}0 & 0 & 0 & 0 & 0 & 0 & 0 & 0 & 0\\
			0 & 1 & 1 & 1 & 1 & 1 & 1 & 0 & 0\\
			0 & 0 & 0 & 0 & 0 & 0 & 0 & 0 & 0
		\end{pmatrix};\:&&\Lambda_{CG_{2}}=\begin{pmatrix}0 & 0 & 0 & 0 & 0 & 0 & 0 & 0 & 0\\
			0 & 0 & 0 & 0 & 0 & 0 & 0 & 0 & 1\\
			0 & 0 & 0 & 0 & 0 & 0 & 0 & 1 & 0
		\end{pmatrix};\\\Lambda_{CG_{3}}=\begin{pmatrix}0 & 0 & 0 & 0 & 0 & 0 & 0 & 0 & 0\\
			0 & 0 & 1 & 1 & 1 & 1 & 1 & 1 & 0\\
			0 & 0 & 0 & 0 & 0 & 0 & 0 & 0 & 0
		\end{pmatrix};\:&&\Lambda_{CG_{4}}=\begin{pmatrix}0 & 1 & 0 & 0 & 0 & 0 & 0 & 0 & 0\\
			1 & 0 & 0 & 0 & 0 & 0 & 0 & 0 & 0\\
			0 & 0 & 0 & 0 & 0 & 0 & 0 & 0 & 1
		\end{pmatrix};
\end{eqnarray*}}

\subsubsection{h=3}\label{s=3,h=3}

\begin{eqnarray*}
	\Lambda_{CG_{1}}&=&\begin{pmatrix}1 & 0 & 0 & 0 & 0 & 0 & 0 & 0 & 0\\
		0 & 1 & 1 & 1 & 1 & 1 & 1 & 1 & 0\\
		0 & 0 & 0 & 0 & 0 & 0 & 0 & 0 & 1
	\end{pmatrix};\\\Lambda_{CG_{2}}&=&\begin{pmatrix}0 & 0 & 0 & 0 & 0 & 0 & 0 & 0 & 0\\
		0 & 1 & 1 & 1 & 1 & 1 & 1 & 1 & 0\\
		0 & 0 & 0 & 0 & 0 & 0 & 0 & 1 & 0
	\end{pmatrix};\\\Lambda_{CG_{3}}&=&\begin{pmatrix}1 & 0 & 0 & 0 & 0 & 0 & 1 & 0 & 0\\
		0 & 1 & 1 & 0 & 1 & 0 & 0 & 0 & 0\\
		0 & 0 & 0 & 0 & 0 & 1 & 0 & 1 & 1
	\end{pmatrix};\\\Lambda_{CG_{4}}&=&\begin{pmatrix}1 & 1 & 0 & 0 & 0 & 0 & 0 & 0 & 0\\
		0 & 0 & 0 & 0 & 1 & 0 & 1 & 1 & 0\\
		0 & 0 & 1 & 0 & 0 & 1 & 0 & 0 & 1
	\end{pmatrix}.
\end{eqnarray*}

\subsection{Some maps and dynamics for stochastic CG results}\label{stochastic_Results}

\subsubsection{Spatial coarse graining}

As before, spatial CG means $h=1$.

\begin{itemize}
	\item $\pi^{(1)}:$ with $\pi^{(1)}$ in the lower level we found seven maps, for example, 
	
	\[
	\Lambda_{CG}=\begin{pmatrix}0 & 0 & 0 & 0 & 0 & 0\\
	0 & 1 & 1 & 1 & 1 & 0\\
	0 & 0 & 0 & 0 & 0 & 0
	\end{pmatrix},
	\]
	which yields in the upper level the following convex combination for the operator related with the first tiling,
	\begin{equation}
	\sigma_0=p_{1}\pi^{(1)}+p_{6}\pi^{(6)},
	\end{equation}
	where $p_{1}, p_{6}\geq0$ and $p_{1}+p_{6}=1$. We also got a convex combination for the operator related with the second tiling $\sigma_1$,
	\begin{equation}
	\label{eq:convex_interaction}
	\sigma_1=q_{1}\mathbb{1}_2+q_{2}\text{swap},
	\end{equation}
	where $q_{1}, q_{2}\geq0$, $q_{1}+q_{2}=1$, and $\mathbb{1}_2$ is the identity permutation. The latter means that, with probability $q_{1}$ the particle will stay in the same cell, and with probability $q_{2}$ the particle will leave the cell.
	\item $\pi^{(2)}:$ in this case only one stochastic evolution is achieved in the upper level,
	\begin{equation}
	\label{eq:CG_2}
	\Lambda_{CG}=\begin{pmatrix}0 & 0 & 0 & 0 & 0 & 0\\
	0 & 0 & 1 & 1 & 1 & 0\\
	0 & 0 & 0 & 0 & 0 & 0
	\end{pmatrix},
	\end{equation}
	which leads to the same evolution expressed in Eq.(\ref{eq:Brownian}), except that now $\pi^{(1)}$ remains the same.
	
	\item $\pi^{(3)}:$ like the result for $\pi^{(2)}$ there is only a single CG map 
	\begin{equation}
	\label{eq:CG_3}
	\Lambda_{CG}=\begin{pmatrix}0 & 0 & 0 & 0 & 0 & 0\\
	0 & 1 & 1 & 1 & 0 & 0\\
	0 & 0 & 0 & 0 & 0 & 0
	\end{pmatrix},
	\end{equation}
	leading to some dynamics in the upper level, where again it is given by Eq.(\ref{eq:Brownian}) for the first operator and swap for the second one.
	
	\item $\pi^{(4)}:$ alternatively from the previous cases, the upper level has a deterministic operator for $\sigma_0$ that is $\pi^{(5)}$, established by the CG map 
	\[
	\Lambda_{CG}=\begin{pmatrix}1 & 0 & 0 & 0 & 1 & 0\\
	0 & 0 & 1 & 1 & 0 & 0\\
	0 & 1 & 0 & 0 & 0 & 1
	\end{pmatrix}.
	\] 
	But the second operator has the format
	\[
	\sigma_1=\frac{1}{2}\mathbb{1}_2+\frac{1}{2}\text{swap},
	\]
	which entails probabilities of $1/2$ for staying or leaving the cell.
	
	\item $\pi^{(5)}:$  in this case, a deterministic evolution for the first operator in the upper level is again achieved, but now the permutation is $\pi^{(4)}$. Coincidentally, with the result achieved for $\pi^{(4)}$ both the CG map and the $\sigma_1$ operator are the same. In fact, by a careful analysis of these permutation operators ($\pi^{(4)}$  and $\pi^{(5)}$) it is possible to see that they are related by a transposition transformation, i.e., $(\pi^{(4)})^{T}=\pi^{(5)}$, the same type of dynamics but for different directions.
	
	\item $\pi^{(6)}:$ finally, for the last permutation operator, there is only one dynamics in the upper level, the same dynamics obtained for $\pi^{(1)}$, according to Eqs. (\ref{eq:Brownian}) and (\ref{eq:convex_interaction}),	achieved by three different CG maps, for instance, 
	\[
	\Lambda_{CG}=\begin{pmatrix}0 & 0 & 0 & 0 & 0 & 0\\
	0 & 1 & 0 & 0 & 1 & 0\\
	0 & 0 & 0 & 0 & 0 & 0
	\end{pmatrix},
	\]
	
\end{itemize}


\subsubsection{Spatial and temporal coarse graining}

Now the results established for $h=2$ are presented.

\begin{itemize}
	\item $\pi^{(1)}:$ the same dynamics for $\pi^{(1)}$ with $h=1$ (see Eqs.(\ref{eq:Brownian}) and (\ref{eq:convex_interaction}) are achieved. However, there are now 63 CG maps doing the same, such as
	\[
	\Lambda_{CG}=\begin{pmatrix}0 & 0 & 0 & 0 & 0 & 0\\
	1 & 1 & 1 & 1 & 1 & 1\\
	0 & 0 & 0 & 0 & 0 & 0
	\end{pmatrix}.
	\]
	\item $\pi^{(2)}:$ the same dynamics achieved in the previous result is kept, but now there is only one map, the one given by Eq.(\ref{eq:CG_3}).  
	
	\item $\pi^{(3)}:$ like the two previous cases, we achieved the same stochastic transition function in the upper level, with Eq.(\ref{eq:CG_2}) as the CG map.
	
	\item $\pi^{(4)}:$ no dynamics is available in the upper level beginning with this deterministic PCA.
	
	\item $\pi^{(5)}:$ as discussed earlier, the dynamics generated by $\pi^{(4)}$ and $\pi^{(5)}$ are quite similar. Given that, we replicated the last result, that is, with no dynamics having been established in the upper level.
	
	\item $\pi^{(6)}:$ now we established three CG maps for only one dynamics in the upper level, which is the same we have seen for $h=1$.
	
\end{itemize}

\bibliographystyle{apsrev4-1}

%

\end{document}